\documentclass[aps,prb,twocolumn,amsmath,amssymb,reprint,groupedaddress]{revtex4-1}
\pdfoutput=1

\usepackage{float}
\usepackage{graphicx}  % needed for figures
\usepackage{xcolor}
\usepackage{dcolumn}   % needed for some tables
\usepackage{bm}        % for math
\usepackage{amssymb}   % for math
\usepackage{epstopdf}
\usepackage{xfrac}
\usepackage{subfigure}
\usepackage{anyfontsize}

\usepackage{amsmath}
\usepackage{amsfonts}
\usepackage{bbm}
\definecolor{darkblue}{rgb}{0.1,0.2,0.6}
\definecolor{darkred}{rgb}{0.8,0.1,0.2}
\usepackage[colorlinks,citecolor=darkblue,linkcolor=darkred,urlcolor=darkblue]{hyperref} 
\usepackage{tikz}
\usepackage{enumerate}
\usepackage{setspace}
\usepackage{url}  % This makes \url work

\numberwithin{equation}{section}

\begin{document}

%Title of paper
\title{Phase diagrams of disordered Weyl semimetals}
\author{Hassan Shapourian}
\author{Taylor L. Hughes}
\affiliation{Department of Physics, University of Illinois at Urbana-Champaign, Urbana Illinois 61801, USA}
\date{\today}

\begin{abstract}
Weyl semimetals are gapless quasi-topological materials with a set of isolated nodal points forming their Fermi surface. They manifest their  quasi-topological character in a series of topological electromagnetic responses including the anomalous Hall effect.
Here we study the effect of disorder on Weyl semimetals while monitoring  both their nodal/semi-metallic and  topological properties through computations of the localization length and the Hall conductivity.
We examine three different  lattice tight-binding models which realize the Weyl semimetal in part of their phase diagram and look for universal features that are common to all of the models,  and interesting distinguishing features of each model.
We present detailed phase diagrams of these models for large system sizes and we find that weak disorder preserves the nodal points up to the diffusive limit, but does affect the Hall conductivity.
We show that the trend of the Hall conductivity is consistent with an effective picture in which disorder causes the Weyl nodes move within the Brillouin zone   along a specific direction that depends deterministically on the properties of the model and the neighboring phases to the Weyl semimetal phase. We also uncover an unusual (non-quantized) anomalous Hall insulator phase which can only exist in the presence of disorder. 
\end{abstract}

\maketitle
%\small
%\singlespacing

\section{Introduction}

The topological classification of symmetry protected fermionic systems has added a new level of complexity to the study of quantum phases of matter. The classification scheme was originally considered for gapped phases~\cite{Schnyder_class,Kitaev_class,Ryu_class}, motivated by the discovery of symmetry protected topological phases with discrete anti-unitary symmetries~\cite{KaneRev,ZhangRev}. Over the past few years there has been a tremendous interest in understanding various topological phases in gapless materials as well,  both experimentally~\cite{PhysRevLett.111.246603,Neupane2014,Liu2014,PhysRevX.5.031013,Xu2015_TaAs,*Xu2015_NbAs,*Zhang2015_ABJ,*Zhang2015,Xu2015_TaP,moll2015}  and theoretically~\cite{NielsenSolidState,Volovik1,*Volovik2,Burkov_Balents1,Burkov_Balents2,Yang_WSM,Wan_arcs,PhysRevLett.107.186806,PhysRevB.86.054504,PhysRevB.86.115133,PhysRevB.86.214514,
PhysRevLett.108.266802,PhysRevLett.108.140405,Delplace,PhysRevB.85.195320,PhysRevB.85.035103,PhysRevB.86.195102,Turner2013293,
Matsuura2013,PhysRevB.88.125105,PhysRevB.88.125110,PhysRevB.89.117101,Zhou2013,PhysRevLett.111.027201,PhysRevB.87.235306,Haldane2014,
Nagaosa2014,PhysRevLett.113.187202,Hosur2013857,PhysRevX.4.031035,Burkov_rev,PhysRevB.89.081407,Ramamurthy,PhysRevB.91.125438,Baum,PhysRevX.5.011029,Huang_dWSM,Lee_Fermiarc} . Aside from graphene in  two dimensions, the Weyl semimetal (WSM) is the most celebrated example of a quasi-topological gapless phase in three dimensions. Its band structure is gapped everywhere in the Brillouin zone (BZ) other than few isolated points, i.e.,  Weyl nodes, where the conduction and valence bands touch to form a doubly-degenerate point. The Weyl nodes are locally stable, topological objects in momentum space~\cite{Volovik1,*Volovik2}, and they each carry a chirality quantum number (Berry phase magnetic charge). Owing to the presence of these nodes, a WSM can display exotic properties such as surface Fermi arcs~\cite{Wan_arcs,PhysRevB.86.195102,PhysRevX.5.031013,Zhang2015,Xu2015_TaP,Lee_Fermiarc}, negative magnetoresistance~\cite{PhysRevLett.111.246603,Zhang2015,Burkov_neg_res}, a magnetic torque anomaly\cite{moll2015}, and an anomalous Hall effect~\cite{PhysRevLett.111.246603,PhysRevLett.113.187202,PhysRevX.4.031035,PhysRevLett.107.186806,Hosur2013857,Burkov_rev}.

Weyl nodes are believed to be robust against perturbations unless discrete translation symmetry or charge conservation symmetry is broken, both of which would allow the nodes to mix with each other and open a gap.
In this paper, we investigate the robustness of WSM phases after breaking translational symmetry via real-space quenched disorder. Stability of a single Weyl node against disorder has been a subject of great effort using continuum limit calculations~\cite{fradkin1,*fradkin2,Burkov_Balents1,PhysRevB.79.045321,PhysRevLett.107.196803,Hosur,Syzranov,PhysRevB.91.035133,Biswas,Ominato,Nandkishore,PhysRevB.92.035203}, as well as numerical simulations~\cite{Huang_13,Sbierski,Sbierski_cr_exp}.  The consensus is that a Weyl node (besides non-perturbative rare region effects~\cite{Nandkishore,PhysRevB.91.035133}) remains a node, with vanishing density of states, up to some finite critical disorder, at which it transforms into a diffusive metal with a finite density of states at the nodal point. 
 
 Unlike these previous studies, the focus of our article is on lattice tight-binding (TB) models with pairs of Weyl nodes. Considering TB models is motivated by the following reasons. First, a single Weyl node cannot appear in isolation in a solid state material according to the Nielsen-Ninomiya no-go theorem~\cite{Nielsen1981219}. Second, some of the gapped regions in the WSM phase can be thought of as  layered topological phases (stacks of Chern insulators) in their own right. The gapped regions lead to their own experimental consequences (anomalous Hall effect, surface states, etc.) and may impact the outcome.  Hence, other observable properties such as Fermi arcs or the anomalous Hall effect (AHE) need to be taken into account in order to determine the nature of a disordered WSM phase.
Measuring topological responses requires the full lattice model to include all non-local (in momentum space) contributions and cannot be done using the low energy effective theory alone.
Third, it is more natural to address the stability of a WSM with reference to its neighboring phases in the full phase diagram of a given lattice model. This is a clear way to identify which phase(s) the WSM transitions to as the system becomes more disordered.

In order to quantify our investigation of the stability of WSM phases we need to keep track of quantities which are special to the WSM and thoroughly characterize it. In other words, one must check simultaneously what disorder does to the topological character of the occupied states as well as the semi-metallic behavior of the Weyl nodes. To monitor the topological properties, we study the AHE by computing the Hall conductivity $\sigma_{xy},$ and to establish the semi-metallic behavior we calculate the localization length. 
We find that the usual expectation that the disorder will mix the Weyl nodes and tend to destroy the WSM phase is not always the case. Indeed there are instances where disorder makes the WSM phase stronger.
There are several interesting recent works~\cite{Tomi_DOS,Pixley_dis_Dirac,Pixley_dis_Dirac2,Liu_Tomi,Bera_DOS} that have also studied TB models of Weyl/Dirac semimetals. In these works they have been interested in critical scaling properties near the metal-insulator transition point evaluated by computing density of states, localization length, and/or the longitudinal conductivity. Our work is markedly different from these studies in the sense that we consider both semi-metallic and topological aspects of disordered WSMs.
A recent, independent study~\cite{CZC} in parallel to our work has also pursued this direction, and has mapped out a phase diagram for a disordered multi-layer Chern insulator model, albeit with a different type of disorder. In that work the phase boundaries are also found to match with the self-consistent Born approximation (SCBA) calculations in the weak disorder limit. In our work, we consider three models, the first of which is similar to the one partially studied in Ref.~\onlinecite{CZC} (where the results overlap they are in complete agreement), and we attempt to recognize the common features among them. 
In explaining our observations, we try to adopt a view based on the generic properties of models (in terms of the distribution of Chern numbers in the BZ) rather than trying to solve the SCBA explicitly for each model. Moreover, in terms of methodology, we compute $\sigma_{xy}$ using the much more efficient Green function approach~\cite{MacKinnon_1980,MacKinnon_1981,MacKinnon_1985} compared to the exact diagonalization method used in Ref.~\onlinecite{CZC}. The Green function method lets us carry out our calculations for large system sizes where exact diagonalization is unwieldy. 

As mentioned, one of the essential goals of our work is to identify which properties and trends of disordered WSM lattice models are universal. For this purpose, we study three different lattice realizations of WSMs. These TB models contain at least one WSM phase in their phase diagrams:  (i) a multi-layer Chern insulator (CI) where layers have Chern number $C=\pm 1$,  (ii) a double WSM model with layers with Chern number $C=2$, and (iii) a 3D, four-band Dirac model tuned near the critical point between the topological insulator (TI) and trivial insulator phases, subject to a Zeeman splitting. (i) and (iii) support Weyl nodes with Berry-monopole charge $\pm 1$ while (ii) is an example of double-WSM having Weyl nodes with Berry-monopole charge $\pm 2$.  In summary, our results show that the Weyl nodes, regardless of their charge and model Hamiltonian, survive at weak disorder and effectively move either towards or away from each other. We find that the movement direction is determined by the sign of a quadratic term in the in-plane momenta normal to the line separating the Weyl nodes. Moreover, we find that a normal insulator (NI) may be transformed into a WSM at finite disorder strength. In this case Weyl nodes nucleate at a finite disorder strength and give rise to a finite AHE response. Inversely, disorder can hybridize the Weyl nodes and hence transform a WSM to a 3D CI (or TI) with a bulk (or surface respectively) AHE.
We show that all these observations are consistent with the SCBA analysis of disordered CIs~\cite{Groth_SCBA}. 
Overall, the effect of the SCBA is to shift  the phase boundaries in each phase diagram as a function of disorder strength. In multi-layer CIs, models (i) and (ii), the WSM region of the phase diagram is maintained up to the diffusive limit; whereas in model (iii) a novel disorder-driven insulating phase with non-zero bulk AHE emerges at the top of WSM region before merging into the diffusive metal. 
In all cases, the AHE is continuous as the WSM region transitions into the diffusive metal (DM) and smoothly decreases to zero on the diffusive side as we reach the Anderson localized regime.  This decreasing trend in the AHE is a direct consequence of diffusive extended bulk states which allow scattering between counter-propagating edge modes on opposite sides through the bulk.

In Sec.~\ref{sec:TB} we start by briefly introducing our methods, next we discuss the phase diagrams of the above models in three subsequent Secs.~\ref{sec:3DCI},~\ref{sec:dWSM}, and~\ref{sec:3DTI}, respectively; finally, we conclude our article in Sec.~\ref{sec:discussion}.

%%%%%%%%%%%%%%%%%%%%%%%%%%%%%%%%%%%%%%%%%%%%%%%%%%%%%

\section{\label{sec:TB} Tight-binding models for the Weyl semimetal}
In this section, we introduce three tight-binding models and study their properties as disorder is added to the system.
Throughout this paper, the disorder is treated as a random on-site energy:
\begin{align}
H_{dis}=  \sum_{\textbf{x}} v_{\textbf{x}} c^\dagger_{\textbf{x}} c_{\textbf{x}},
\end{align}
where $c_{\textbf{x}}$ and $c^\dagger_{\textbf{x}}$ are electron destruction and creation operators (and may have extra spin/orbital indices) and $v_{\textbf{x}}$ is a random number uniformly distributed in the range $[-W/2,W/2]$ where $W$ is the disorder strength. For each model we consider we will present a phase diagram where each phase is characterized by the Hall conductivity $\sigma_{xy}$ and the localization length. %, or the longitudinal conductance $\sigma_{xx}$.
 In order to calculate the localization length, we solve the standard recursive equation by viewing the system as a stack of layers along the $z$ direction~\cite{MacKinnon_PRL,MacKinnon_TM2,Pichard}. The localization length is determined by $\xi=1/\kappa$  in terms of the smallest positive Lyapunov exponent $\kappa$ for a quasi-one-dimensional system of  cross section $L_x\times L_y=L\times L$ and length $L_z$ using the resolvent (Green) function method which does not require any additional condition on the invertibility of the inter-layer hopping matrix. In our calculations, we take disorder ensemble averages for system sizes of $L=8,10,12,14$ and $L_z=10^4$. 
 We should note that the systems are quasi-one dimensional ($L_z \gg L$) which means that they are always localized or 
 equivalently the transfer matrix always decays exponentially with $L_z$, i.e. $T \propto \exp(-2L_z/ \xi)$.
 However, one can still distinguish metallic phases from insulating phases by looking at how $\xi$ behaves as the cross section is made larger. Let the normalized localization length be $\Lambda=\xi/L$. Therefore, in a (semi-) metallic phase $\Lambda$ is proportional to L which implies $\xi \propto L^2$ (or the number of conducting channels), whereas in an insulating phase $\Lambda$ is decreasing as 1/L which means $\xi$ is almost constant. Any critical point is characterized by a scale invariant $\Lambda$ where all curves near this point fall onto a single-parameter scaling ansatz.
 We use these facts to determine the boundaries between various phases in our phase diagrams. We impose the periodic boundary conditions across each slice along $x$ and $y$ directions.

For a clean WSM with two nodes at $\pm k_0$ along the $z$-direction, the Hall \emph{conductivity} in the normal plane is known to be 
\begin{align} \label{eq:AHE}
\sigma_{xy}= \frac{1}{L_z} \sum_{k_z} \sigma_{xy}^{2D} (k_z) = \frac{e^2 k_0}{\pi h}
\end{align}
where $\sigma_{xy}^{2D} (k_z)$ is the Hall \emph{conductance} of a 2D layer at $k_z$ in the 3D BZ~\cite{Burkov_Balents2}. In the rest of this paper, we set $e=h=1$.
To compute $\sigma_{xy}$  in a disordered media, we use the Green function method developed in Refs.~\onlinecite{MacKinnon_1980,MacKinnon_1981,MacKinnon_1985}. 
The Green function approach is very efficient for 3D calculations since in this method the 3D sample is divided into 2D slices normal to a given direction (e.g., the $x$-direction) and the response function is computed iteratively as more layers are added through a set of recursion relations. 
The idea is to write the Hall conductivity as
\begin{align}\label{eq:AHE}
\sigma_{xy} =&\frac{4}{L_x L_y L_z} \text{Tr} {\Big[} \gamma^2 \sum_{i,j}^{L_x} \boldsymbol{G}^+_{ij} \boldsymbol{y} \boldsymbol{G}^-_{ji} x_i \nonumber \\
&- i \frac{\gamma}{2} \sum_{i}^{L_x} (\boldsymbol{G}^+_{ii}-\boldsymbol{G}^-_{ii}) x_i \boldsymbol{y} {\Big]}
\end{align}
for a 3D sample with $L_x$ slices each containing $L_y L_z$ sites. Here $\boldsymbol{G}^{\pm}_{ij}$ is the retarded (advanced) Green's function connecting the $i$-th and $j$-th layers, and the position operators are denoted by ${\bf x}$ and ${\bf y}$. $\boldsymbol{G}^{\pm}_{ij}$ is computed at $Z=\mu+i \gamma$ where $\mu$ is the chemical potential and the phenomenological lifetime parameter $\gamma$ is introduced to avoid singular behavior. Ultimately the limit $\gamma\to 0$ is assumed. $\boldsymbol{G}^{\pm (n+1)}_{ij}$ is evaluated at the $(n+1)$-th step (for $n+1$ layers) from the Green function $\boldsymbol{G}^{\pm (n)}_{ij}$  (for $n$ layers) by updating the self-energy with the contribution from the $(n+1)$-th layer. 
The efficiency of this approach is of particular importance here as we need to consider rather large system sizes to get accurate results for $\sigma_{xy}$. Other methods such as the real-space formula for the Chern number based on projected position operators~\cite{Prodan_1,Prodan_2} (which was also used in Ref.~\onlinecite{CZC}) involves projection operators onto the occupied states, and hence requires a full diagonalization the 3D Hamiltonian.  The diagonalization process is memory intensive and very time consuming leading to long processing times which scale as $O((L_xL_yL_z)^3)$ compared to $O( L_x (L_yL_z)^3)$ for the Green function method.

To calculate $\sigma_{xy}$ with high accuracy, we choose the system size by increasing the lengths in all directions until $\sigma_{xy}$ converges (Eq.~(\ref{eq:AHE})) and further increases do not modify the result significantly. Our observation is that one usually needs to go to larger lengths ($> 30$) in the $x$ and $y$ directions while rather small lengths ($\sim10$) in the longitudinal $z$-direction often work well. This can be understood by noting that the system needs to be large enough in the $xy$ plane to yield enough k-points for precise evaluation of the Chern number, and should be long enough in the $z$-direction to give a sufficient number of layers between the Weyl nodes. For calculating $\sigma_{xy}$ in this paper, the boundary conditions (BC) across each layer are chosen to be periodic along the $z$ direction and open along $y$ direction. We note that while $\sigma_{xy}$ in the WSM and NI phases does not depend on BC in $z$ direction, it does depend on the BCs in case of 3D TIs (studied in Sec.~\ref{sec:3DTI}) due to the extra half-integer surface Hall conductance arising from the gapped Dirac surface states.

%%%%%%%%%%%%%%%%%%%%%%%%%%%%%%%%%%%%%%%%%%%%%%%%%
\subsection{\label{sec:3DCI} Stack of Chern insulators C=1}
As the simplest model of the WSM, we consider the two-band Hamiltonian of multi-layer CI model~\cite{Burkov_Balents2,Yang_WSM}
\begin{align} \label{eq:3DChern}
H=& \frac{1}{2} \sum_{\substack{\textbf{x}\\ s=1,2}} {\Big[} c_{\textbf{x}+\textbf{a}_s}^\dagger (i t \sigma_s - r \sigma_3) c_\textbf{x} +\text{h.c.} {\Big]}
 + m \sum_\textbf{x} c_\textbf{x}^\dagger \sigma_3 c_\textbf{x} \nonumber \\
&- \frac{t_3}{2} \sum_{\textbf{x}} {\Big[} c_{\textbf{x}+\textbf{a}_3}^\dagger \sigma_3 c_\textbf{x} +\text{h.c.} {\Big]}
\end{align}
where $c,c^\dagger$ denote two component fermion operators, $\sigma_i$ are Pauli matrices representing spin, and $\textbf{a}_s$ are cubic lattice unit vectors $|\textbf{a}_s|=1$. The first two sums describe a stack of CI layers coupled to their neighboring layers through the third sum with a tunneling amplitude $t_3$.
Fourier transforming to momentum space, the Bloch Hamiltonian is given by
\begin{align} \label{eq:3DCI_p}
h(\textbf{k})=&  t \sin k_x\ \sigma_1 +t \sin k_y\ \sigma_2 \nonumber \\
&+  ( m-r \cos k_x-r \cos k_y -t_3 \cos k_z) \sigma_3.
\end{align}
We fix the parameters $r=t=1,$ and focus on the regime $0< t_3 \leq m.$ The phase diagram is symmetric under $m \to -m$ up to flipping the sign of $\sigma_{xy}$.
In the limit $m \to \infty$, this model realizes an NI. As $m$ is tuned down to zero, it supports two qualitatively distinct WSM phases:  `WS~I' for $m<t_3$ and `WS~II' for $|m-2r|< t_3$. The WS~I phase contains two pairs of Weyl nodes at $(0,\pi,\pm k_0)$ and $(\pi,0,\pm k_0)$ yielding $\sigma_{xy}=1-2k_0/\pi$ in which $\cos^{-1}m/t_3$. The WS~II phase hosts a pair of Weyl nodes at $(0,0,\pm k_0)$ and has $\sigma_{xy}=k_0/\pi$ where $k_0=\cos^{-1}(m-2r)/t_3$. 
The intermediate region between these two WSM phases is described by a 3D CI (i.e., a 3D T-breaking weak topological insulator) where $\sigma_{xy}=1$.

\begin{figure}
\includegraphics[scale=0.08]{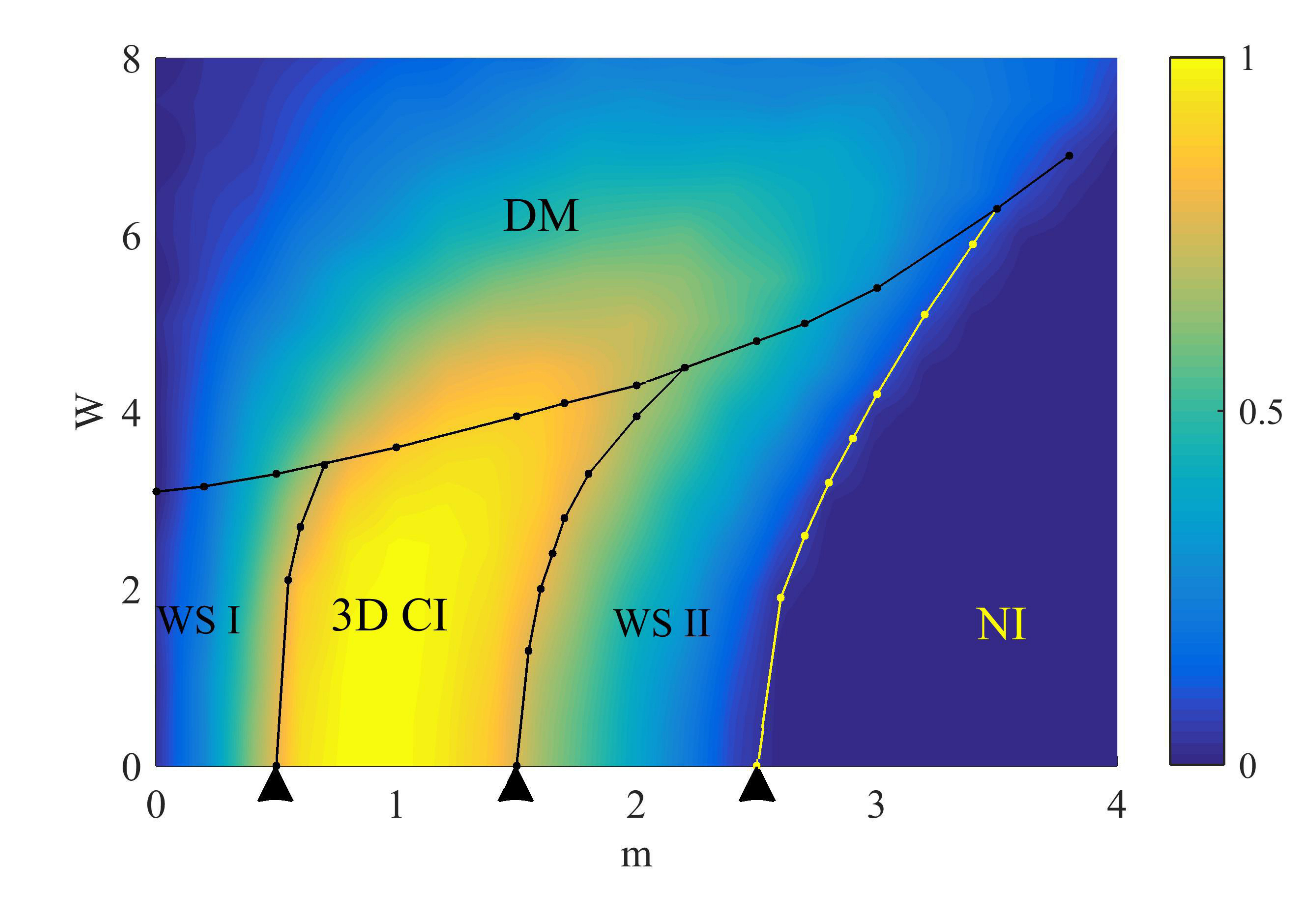}
\includegraphics[scale=0.08]{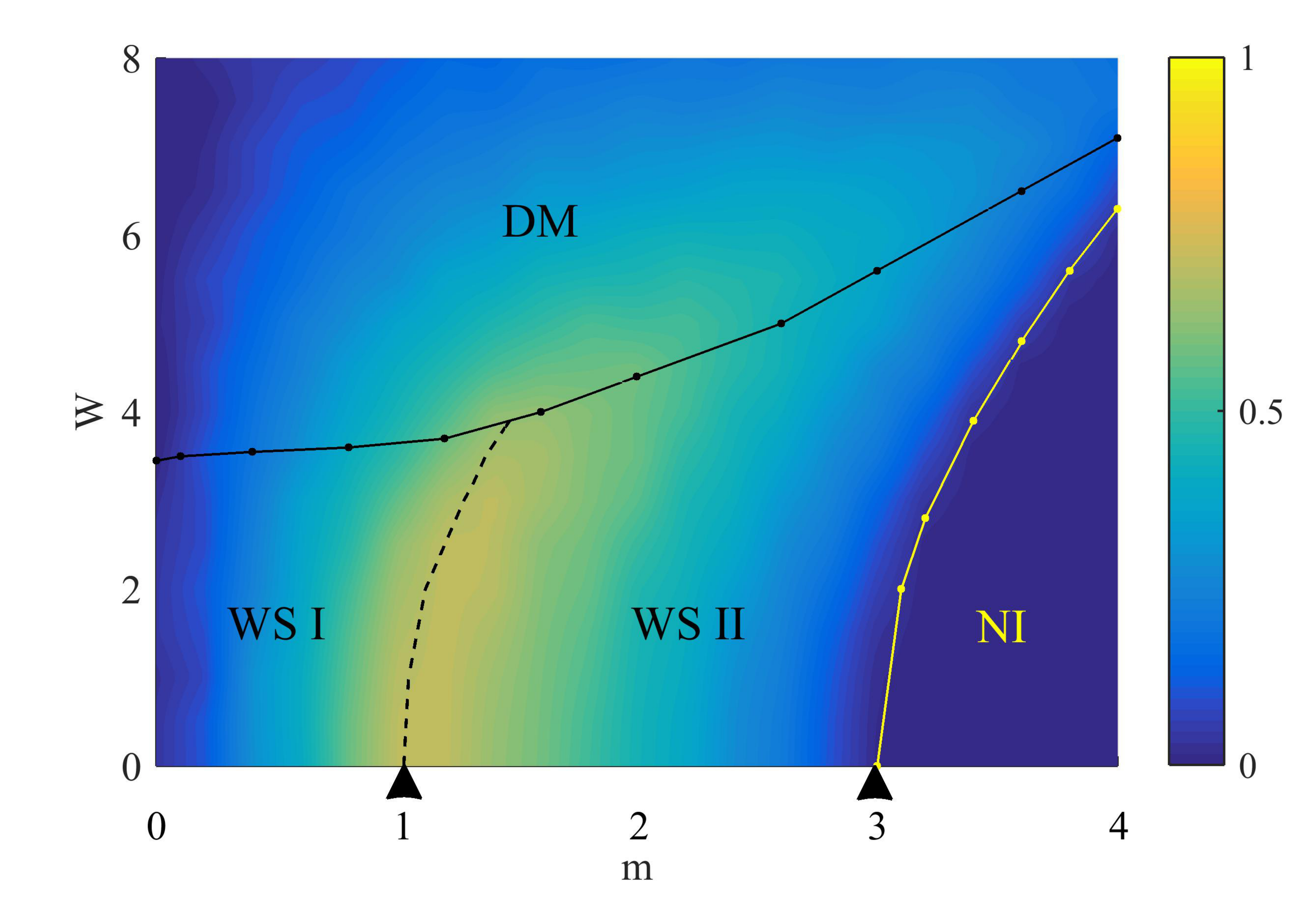}
\caption{\label{fig:AHE_CI}(Color online)  Phase diagram of the layered Chern insulator model given in Eq.~(\ref{eq:3DChern}). The color map represents $\sigma_{xy}$ and the solid lines representing phase boundaries are determined by the localization length. We show the phase diagram for two different values of $t_3$: $t_3=0.5$ (top) and $t_3=1$ (bottom). The system size is $30 \times 30 \times 12$. Black triangles on the $x$-axis mark the clean phase boundaries. The dashed line in the lower panel connects maxima of $\sigma_{xy}$ as  a guide for the eye. We note that there is no difference in meaning between the solid black and solid yellow lines, the color difference is just to add contrast.}
\end{figure}

\begin{figure}
\includegraphics[scale=0.64]{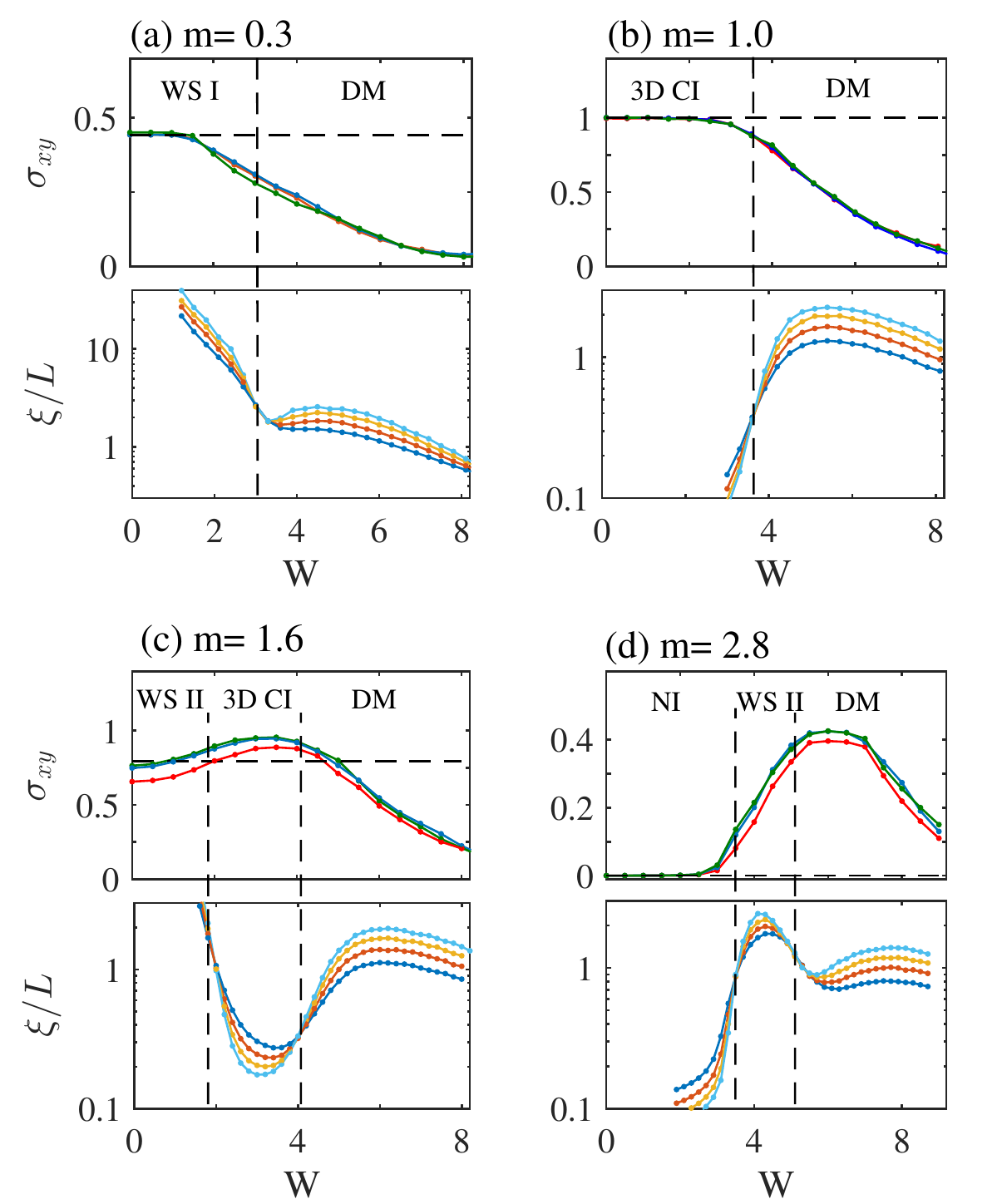}
\caption{\label{fig:CI_comp} The Hall conductivity and localization length versus disorder strength for layered Chern insulator model with  $t_3=0.5$. In the lower panel  of each subfigure, the colors show different cross-section sizes $L=8$(blue), $10$(red), $12$(orange), and $14$(cyan). The dashed vertical lines represent phase boundaries, the dashed horizontal lines indicate the  value of $\sigma_{xy}$ in the clean limit using the analytical expressions explained in the text, and the system size for $\sigma_{xy}$ are $ 30 \times 30 \times 12$ (red), $70 \times 70 \times 16$ (blue), and $90 \times 90 \times 20$ (green).}
\end{figure}

\begin{figure}
\includegraphics[scale=1]{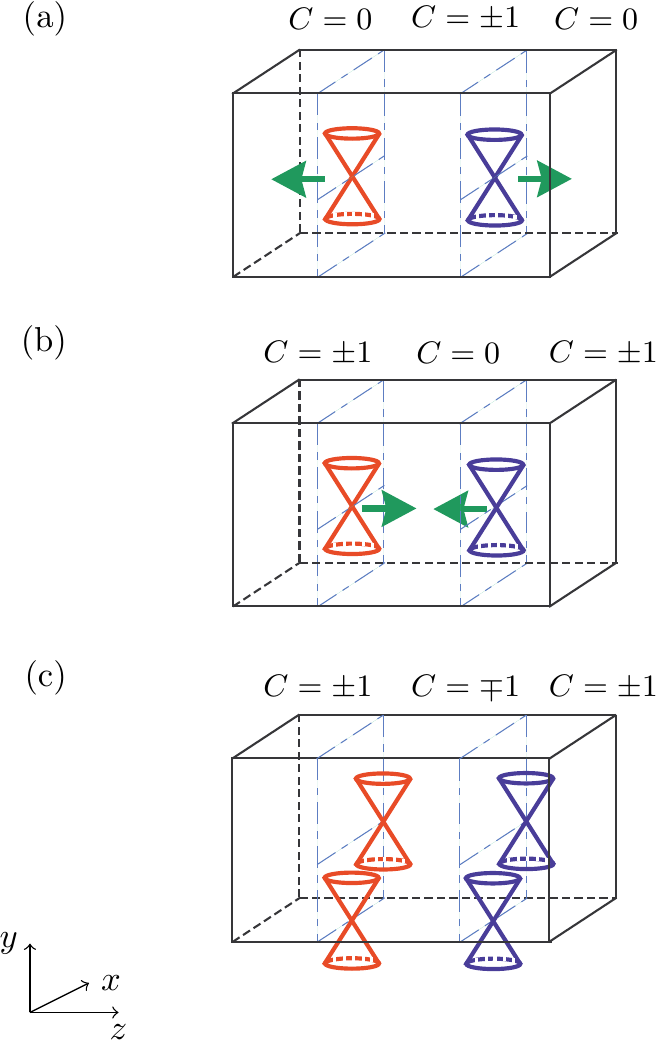}
\caption{\label{fig:motion}  The effective shifting directions of the Weyl nodes inside BZ for the layered CI model as the disorder strength is increased (but below the diffusive metal regime). The Chern number of the momentum slices in each section of the BZ is shown above them. Weyl cones of different colors represent opposite chiralities. This motion is determined by the SCBA calculation and is consistent with the numerical results.  For subfigure (c) the Weyl nodes do not move.}
\end{figure}

Figure~\ref{fig:AHE_CI} presents the phase diagrams in the $(m,W)$ plane for two different values of the interlayer hopping $t_3=0.5$ (top) and $t_3=1.0$ (bottom). In the former case, two WSM phases (denoted by WS~I and WS~II) are separated by a gapped 3D CI phase (i.e., a 3D T-breaking weak TI phase). As the colormap shows, in both cases $\sigma_{xy}$ can be used to distinguish various phases even at finite disorder. The solid lines indicate the phase boundaries determined by the scale invariant points where the localization length $\xi$ is size independent. 
Comparisons of $\sigma_{xy}$ and $\xi$ for few values of $m$ are shown in Fig.~\ref{fig:CI_comp}. The top panels in  Fig.~\ref{fig:CI_comp} also compare $\sigma_{xy}$ for different sizes and confirm that the finite size corrections are rather small once we have reached a certain size.

There are several interesting features in these plots. As the system becomes diffusive, denoted by the label diffusive metal (DM), $\sigma_{xy}$ starts to decrease and eventually vanishes for ultra-strong disorder ($W>16$) in the Anderson localized phase. 
In addition, the WSM phases tend to broaden as the disorder becomes stronger. The critical points in the clean limit separating WS~I-3D~CI-WS~II-NI phases (Fig.~\ref{fig:AHE_CI}(top)) or  WS~I-WS~II-NI  phases (Fig.~\ref{fig:AHE_CI}(bottom)) are moved to the right as disorder strength increases. In particular, $\sigma_{xy}$ increases when disorder \emph{increases} in the WS~II phase (Fig.~\ref{fig:CI_comp}(c)). 

This behavior, and many of the results that follow, can be understood by thinking about the clean limit subject to weak disorder. Since our localization length data shows that, starting in the WSM phase, there is no critical behavior until the DM phase is reached, we might hope that our weak disorder understanding will apply for a wide-range of disorder strengths; indeed by comparison with our numerical results the application appears successful. Now, 
based on the picture of a WSM as CI layers in momentum space\cite{Burkov_Balents2,Yang_WSM,Ramamurthy}, then as far as the low-energy Hamiltonian is concerned, the 3D BZ can be viewed as layers of CI with a varying mass parameter 
\begin{align}
h(\textbf{k})\approx t (k_x \sigma_1 + k_y \sigma_2) + (M(k_z)+ \frac{m_\perp}{2}k_\perp^2 ) \sigma_3
\end{align}
where $k_\perp^2=k_x^2+k_y^2$ is the in-plane momentum, $M(k_z)=m-2r-t_3 \cos k_z,$  and $m_\perp=r$. Hence, in the direction of separation between the nodes the BZ can (usually) be divided into topological layers $M(k_z)< 0$ of non-zero Chern number and trivial layers $M(k_z)>0$ of zero Chern number. The planes containing the Weyl nodes at $k_z=\pm k_0$ we will call the `Weyl planes' where $M(\pm k_0)=0$. 

With this set up, let us consider the effects of disorder. Consider two Weyl planes in the BZ where either the layers inside (Fig.~\ref{fig:motion}(a)) or outside (Fig.~\ref{fig:motion}(b)) the two planes are topological and the complementary region is trivial.   Numerically, our results can be consistently interpreted if the effect of disorder moves the Weyl planes in the direction such that the number of topological layers increases (Fig.~\ref{fig:motion}). This means that insulating layers near the Weyl planes on the trivial side become topological with non-zero Chern number. This effect can happen since the disorder can be thought to renormalize the mass parameter down to negative values and hence inverting the bands. This type of behavior has also been shown for a low-energy model CI using the SCBA and including the quadratic (in momentum) mass terms~\cite{Groth_SCBA}. Using only the Born approximation (i.e., not self-consistent), they found the mass renormalization $\delta M$ to be~\cite{Groth_SCBA}
\begin{align} \label{eq:CImass_ren}
\delta M = - \frac{W^2 }{48 \pi m_\perp} \ln {\Big|} \frac{\pi^2  m_\perp}{2 M }{\Big|}.
\end{align}
One important consequence of this result is that sign of $\delta M$ depends on the sign of quadratic term $m_\perp k_\perp^2$ as follows: if $m_\perp>0$, then $\delta M<0$ and vice versa.

For our model, $m_\perp=r$ is always positive and hence the mass renormalization is always negative.
This is then a way to understand the underlying process leading to an enhanced AHE response in the WS~II phase as a function of disorder strength.
Moreover, we observe that the insulating phase NI in the vicinity of WS~II transitions into a WSM at a finite disorder before transforming to the DM (Fig.~\ref{fig:CI_comp}(c)). This can also be attributed to the negative mass renormalization which makes the trivial layers near the origin topological and lets Weyl nodes nucleate at the interface between newly formed topological layers and trivial ones. 

However, a different situation occurs in the WS~I phase; i.e.,~$\sigma_{xy}$ starts as a plateau with initially increasing disorder, and then decreases as disorder is further increased (Fig.~\ref{fig:CI_comp}(a)).
In this phase, there are two pairs of Weyl nodes with the same chirality in the two planes at $k_z= \pm k_0$ allowing for the Chern number to change by two as the Weyl planes are crossed in momentum space (Fig.~\ref{fig:motion}(c)). Hence, the layers between Weyl planes $|k_z| < k_0$ have $C=-1$ and the others $|k_z|> k_0$ have $C=1$. Hence, all of the layers are topological in the clean limit. This is different than the usual form seen in the WS~II phase. 

Interestingly, unlike WS~II, we find that the SCBA mass renormalization for these nodes is zero, and hence, the trend of $\sigma_{xy}$ in this case cannot be immediately determined as a perturbative effect as it can be in WS~II. This is one reason why we might expect (and as we do see numerically) that $\sigma_{xy}$ should remain constant at weak disorder.
To help understand this behavior beyond perturbation theory, we numerically calculated the phase diagram of the 2D CI in Appendix A. By applying the results of this phase diagram  we expect that rather weak disorder can localize the edge modes of CI layers on both sides of Weyl planes, which have opposite Chern number, and will nucleate a region with $C=0$ beginning in each Weyl plane. One might understand this by considering that, since nearby momentum slices on the opposite sides of the Weyl plane have opposite Chern number, they can couple at weak disorder and annihilate. In fact,  since, the reduction of Chern number on both sides would be symmetric, we would again expect that we will not find any change in $\sigma_{xy}$ initially. However, we note that $k_0<\pi/2$ which means the region with $C=1$ ($|k_z|> k_0$) is wider than the region with $C=-1$ ($|k_z|< k_0$ ). Thus, as we continue increasing the disorder, the central region$|k_z|< k_0$  becomes entirely trivial first, while part of the $|k_z|> k_0$ region remains topological. So, a further increase of disorder results in making these layers trivial and consequently reduces $\sigma_{xy}.$

\subsection{\label{sec:dWSM} Stack of Chern insulators C=2}

A simple model of a double Weyl semimetal (dWSM) can be constructed by stacking layers of CIs with Chern number two,
\begin{align} \label{eq:DWS}
H=& \frac{1}{2} \sum_{\substack{\textbf{x}\\ s=1,2}} {\Big[} c_{\textbf{x}+\textbf{a}_s}^\dagger (t \sigma_1 -  r \sigma_3) c_\textbf{x} +\text{h.c.} {\Big]}
 + m \sum_\textbf{x} c_\textbf{x}^\dagger \sigma_3 c_\textbf{x} \nonumber \\
& + \frac{t'}{2} \sum_{\substack{\textbf{x}\\ s=1,2}} {\Big[} c_{\textbf{x}+\textbf{a}'_+}^\dagger  \sigma_2  c_\textbf{x} - c_{\textbf{x}+\textbf{a}'_-}^\dagger  \sigma_2  c_\textbf{x} +\text{h.c.} {\Big]} \nonumber \\
&- \frac{t_3}{2} \sum_{\textbf{x}} {\Big[} c_{\textbf{x}+\textbf{a}_3}^\dagger \sigma_3 c_\textbf{x} +\text{h.c.} {\Big]}
\end{align}
where $\textbf{a}_s$ are unit vectors for the cubic lattice, and $\textbf{a}'_{\pm}=\textbf{a}_1\pm \textbf{a}_2$ connect next nearest neighbors in the $xy$-plane. In the clean limit, the Bloch Hamiltonian can be written as
\begin{align}
h(\textbf{k})=&  t (\cos k_x-\cos k_y)\ \sigma_1 +2 t' \sin k_x\sin k_y\ \sigma_2 \nonumber \\
&+  ( m-r \cos k_x -r \cos k_y -t_3 \cos k_z) \sigma_3.
\end{align}
When $|m-2r|<t_3$, this model yields Weyl nodes at $(0,0,\pm k_0)$ each with a Berry-monopole charge of $\pm 2$ where $k_0=\cos^{-1} (m-2r)/t_3$.  The Hall conductivity is given by $\sigma_{xy}=2k_0/\pi$. 

\begin{figure}
\includegraphics[scale=0.08]{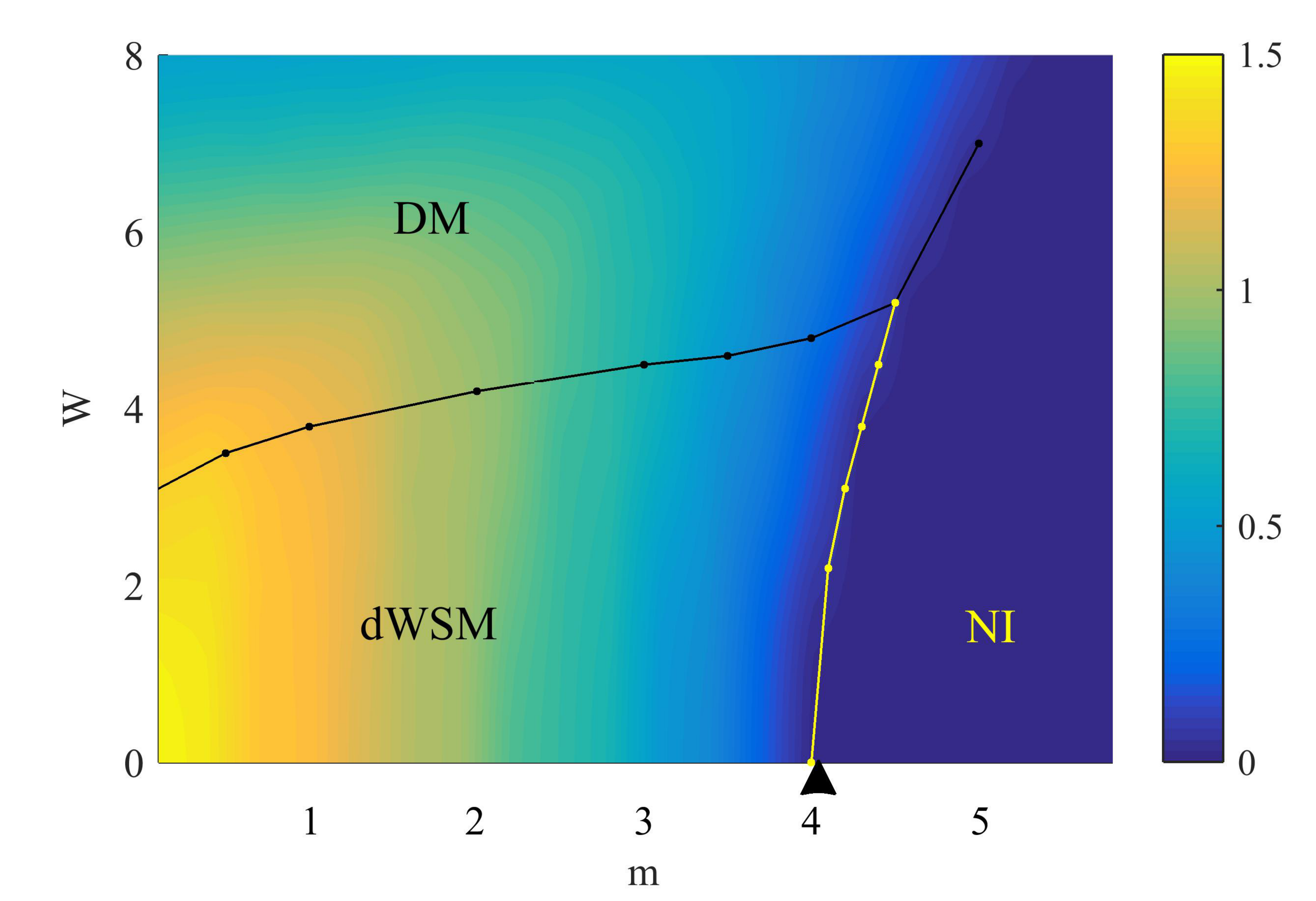}
\caption{\label{fig:AHE_DWS} (Color online)  Phase diagram of the dWSM model given by Eq.~(\ref{eq:DWS}). The color map represents $\sigma_{xy},$ and the solid lines are phase boundaries determined by the localization length. The black triangle on the $m$-axis marks the clean phase boundary.  The system size is $30 \times 30 \times 14$. We note that there is no difference in meaning between the solid black and solid yellow lines, the color difference is just to add contrast.}
\end{figure}

The phase diagram for $t_3=2r$ is shown in Fig.~\ref{fig:AHE_DWS}. It essentially follows the same properties of the multi-layer CI model in the previous section with $C=1$. Similar to that model, in the diffusive limit, the Hall conductivity decreases monotonically to zero.
Moreover, before reaching the diffusive limit we can understand the structure of the phase diagram using the Born-approximation arguments from the previous section. The clean Hamiltonian for a layer at $k_z,$ and near the origin of the in-plane momenta $\textbf{k}_\perp=(k_x,k_y),$ reads
\begin{align}
h(\textbf{k})\approx& -\frac{t}{2} (k_x^2-k_y^2) \sigma_1 + 2 t' k_x k_y \sigma_2 \nonumber \\
&+ (M(k_z)+ \frac{m_\perp}{2}k_\perp^2 ) \sigma_3,
\end{align}
in which $M(k_z)=m-2r-t_3 \cos k_z$ determines trivial layers $M(k_z)>0$ from topological ones $M(k_z)<0$.
Since the phase boundary of the dWSM moves further to the right (larger $m$) then we need to understand the instability of trivial layers near the Weyl nodes,  $M(\pm k_0)=0$, to transition into the topological phase with Chern number two. Nicely, this can again be explained in terms of mass renormalization towards the topological phase similar to the $C=1$ case~\cite{Groth_SCBA}. 
For $t=2t'$, the SCBA, in the Born approximation limit, can be analytically derived to be
\begin{align} \label{eq:DWS_ren}
\delta M = -\frac{ W^2}{ 96 \pi} \frac{ m_\perp}{ m_\perp^2+t^2} \ln \left[\frac{\pi^4(m_\perp^2+t^2)}{4 M^2} \right].
\end{align}
This implies that the Weyl nodes move away from one another towards the BZ boundary so that $\sigma_{xy}$ will  increase (similar to Fig.~\ref{fig:motion}(a)).

Let us now comment on the similarities between the WSMs realized by multi-layer CI models. In terms of the Hall conductivity, they share two common properties: (i) in the weak disorder limit the Weyl nodes are pushed toward the region with zero Chern numbers (see Fig.~\ref{fig:motion}), hence expanding the topological portion of the  BZ and (ii) in the diffusive limit the AHE is weakened as the disorder is made stronger. 
It is worth noting that the former property is specific to the multi-layer CI systems where the sign of quadratic term $r k_\perp^2$ is independent of the position $k_0$ of the Weyl nodes. The model in the next section that is constructed from a 3D Dirac model, will not always show this behavior in the entire phase diagram since the sign of quadratic term depends on the nodal momenta. Given our localization length data, we have not noticed any drastic differences between the dWSM and the ordinary WSM in terms of their stability against disorder or their transitioning to the DM phase at finite disorder. This is in contrast with the conclusions of Ref.~\onlinecite{Bera_DOS}, though we are not sure if the model used in that work should be directly comparable to ours. If they cannot be directly compared then perhaps this would resolve the discrepancy. 

In terms of the longitudinal conductivity, our calculation of localization length suggests that the WSM stays semi-metallic at weak disorder and subsequently becomes a DM for sufficiently strong disorder. In other words, it never becomes insulating in this process. As we will see in the next section, the WSM obtained by breaking the time-reversal symmetry in the 3D Dirac system does not follow this trend and transitions to an intermediate insulating phase along the way transitioning into a DM.

\subsection{\label{sec:3DTI} 3D Dirac semimetal subject to a Zeeman field}

Consider the four-band model of a time-reversal invariant $\mathbb{Z}_2$ TI~\cite{Wilson_3DTI,Qi_3DTI}
\begin{align} \label{eq:3DTI}
H=& \frac{1}{2} \sum_{\substack{\textbf{x}\\ s=1,2,3}} {\Big[} c_{\textbf{x}+\textbf{a}_s}^\dagger (i t \alpha_s - r \beta) c_\textbf{x} +\text{h.c.} {\Big]} \nonumber \\
& + (m+3r) \sum_\textbf{x} c_\textbf{x}^\dagger \beta c_\textbf{x},
\end{align}
where the Dirac matrices are given by
\begin{align*}
\alpha_s&= \tau_1\otimes \sigma_s=\left(\begin{array}{cc}
0 & \sigma_s \\ \sigma_s & 0
\end{array} \right), 
&\beta= \tau_3\otimes 1=\left(\begin{array}{cc}
\mathbbm{1} & 0 \\ 0 & -\mathbbm{1}
\end{array} \right).
\end{align*}
In this convention the $\sigma$ and $\tau$ matrices act on the spin and orbital degrees of freedom respectively. We set the parameters at $r=t=1$ from now on. 

The Bloch Hamiltonian in the clean limit is:
\begin{align*}
h(\textbf{k})= \sum_{s=1,2,3}{\Big[} t \alpha_s \sin k_s - r\beta \cos k_s {\Big]}+ (m+3r)\beta .
\end{align*}
There are two important symmetries of this model: (i) time-reversal symmetry (TRS) with the operator ${\cal T} = i \sigma_2 {\cal K}$ such that $\sigma_2 h(\textbf{k})\sigma_2= h^\ast(-\textbf{k})$; (ii) inversion symmetry (IS), represented by ${\cal I}= \tau_3{\cal P}$, such that $\tau_3 h(\textbf{k})\tau_3=h(-\textbf{k})$. This model realizes a strong TI (STI) phase over the mass range $-2r<m<0,$ and there is a transition from STI to an NI at $m=0$. The critical point at $m=0$ is described by a gapless 3D Dirac Hamiltonian with four bands touching at a single point. If we add a TRS or IS breaking term then the critical point broadens into an intermediate semi-metallic region, i.e.  a Weyl semimetal. In this case the fermions with opposite handedness are shifted away from the $\Gamma$-point and form Weyl nodes at different points inside the BZ.
 
 One such term we could add to break TRS is a Zeeman splitting field:
\begin{align}
H_{b_s}= b_s \sum_\textbf{x} c_\textbf{x}^\dagger \tau_0 \sigma_s c_\textbf{x}
\end{align}
where $0<b_s < t.$ Depending on $s=1,2,3$ this term splits the Dirac node into two Weyl points which are shifted from the origin in opposite directions ($\pm k_0$) along the $s$-axis in momentum space. The amount of shift is given by
\begin{align} \label{eq:4b_k0}
\cos k_0 = \frac{t^2+(m+t)^2-b_s^2}{2t(m+t)}.
\end{align}
Therefore, in the presence of this term, a WSM phase forms in the range $|m|\leq b_s$ between the STI and NI phases. From now on, let the Zeeman field be in the $z$-direction with $b_3\neq 0$.
From the NI side ($m> b_3$), as $m$ is decreased the two Weyl nodes start to nucleate at the origin when $m=b_3$ and move apart from each other towards the BZ boundary until  $m_c=\sqrt{t^2-b_3^2}-t <0$ where they reach their maximum distance. Making $m$ smaller results in bringing the nodes closer and they finally annihilate each other at the origin once $m=-b_3$ is reached. After annihilation the STI phase is formed.   

The spectrum near a Weyl point at $(0,0, k_0)$ is given by
\begin{align} \label{eq:Dirac_eff}
h(\textbf{k})= v_\perp (k_x \eta_1+k_y \eta_2) + (v_z k_z + \frac{m_\perp}{2} k_\perp^2) \eta_3,
\end{align}
where $\eta_i$ are Pauli matrices in the projected space of the two touching energy bands, the in-plane momentum is $k_\perp^2=k_x^2+k_y^2$, and the other coefficients are 
\begin{align}
\label{eq:quad}
v_\perp=m_\perp &=\frac{m+t- t \cos k_0}{b_3} t \\
v_z &=\frac{m+t}{b_3} t \sin k_0.
\end{align}
Note that the chiralities of Weyl nodes at the two points $(0,0,\pm k_0)$ are opposite to each other since only the third component of velocity  flips sign $v_z(-k_0)=-v_z(k_0)$.

\begin{figure}
\includegraphics[scale=0.08]{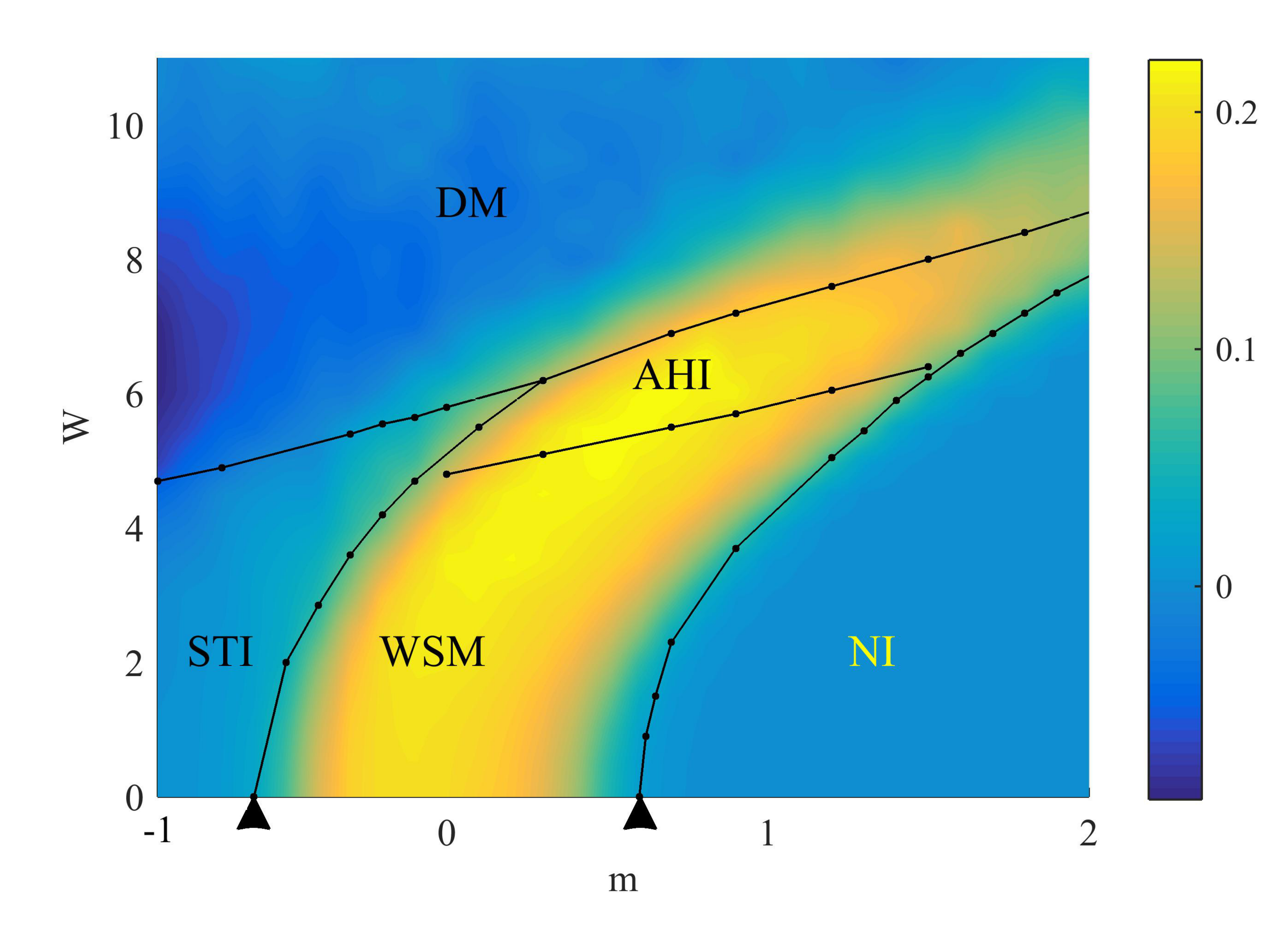}
\caption{\label{fig:AHE_fourband}(Color online)
The phase diagram of a WSM generated by broken TRS in a 3D Dirac model given by Eq.~(\ref{eq:3DTI}). The color map represents $\sigma_{xy}$ and solid lines are phase boundaries determined by localization length.  The AHI phase is contained in a closed region of the phase diagram and the NI-AHI and AHI-DM boundaries eventually intersect at $(m,W)=(2.6,9.5)$ (this lies outside the graph on the top right). The black triangles on the $m$-axis  mark the clean phase boundaries. $b_3=0.6$ and the system size is $30 \times 30 \times 10$. We note that the STI phase would have a quantized surface Hall conductivity if we had chosen open boundaries in the $z$-direction instead of periodic. We checked that this was the case, but did not include both diagrams.}
\end{figure}

\begin{figure}
\includegraphics[scale=0.64]{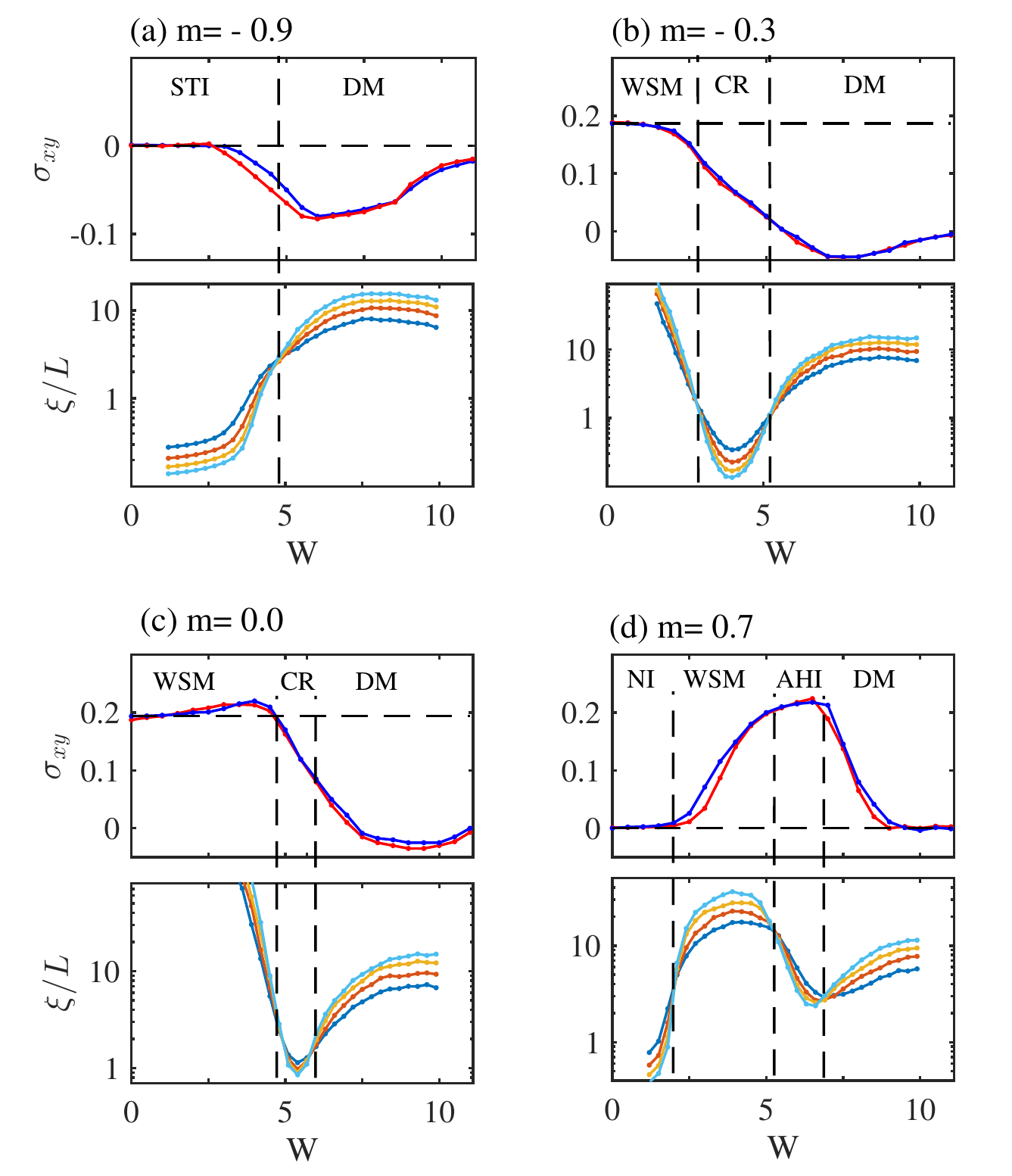}
\caption{\label{fig:4b_comp}
The Hall conductivity and localization length versus disorder strength for  the WSM generated by broken TRS in a 3D Dirac model  with  $b_3=0.6$. The dashed vertical lines represent phase boundaries and the dashed horizontal lines indicate the  value of $\sigma_{xy}$ in the clean limit using Eq.~(\ref{eq:4b_k0}).  For $\sigma_{xy}$, the system size is $30 \times 30 \times 10$ (red) and $50 \times 50 \times 16$ (blue). For $\xi/L$, the colors show different cross-section sizes $L=6$(blue), $8$(red), $10$(orange), and $12$(cyan). We note that in subfigure (a) the STI phase would have a quantized surface Hall conductivity if we had chosen open boundaries in the $z$-direction instead of periodic. We checked that this was the case, but did not include both diagrams. Also, CR in panel (b) and (c) refers to the critical STI phase near the critical line between STI and AHI phases, where the non-zero bulk $\sigma_{xy}$ could be due to finite-size effects (i.e.~$\sigma_{xy}$ varies smoothly from AHI to STI). }
\end{figure}

\begin{figure}
\includegraphics[scale=0.72]{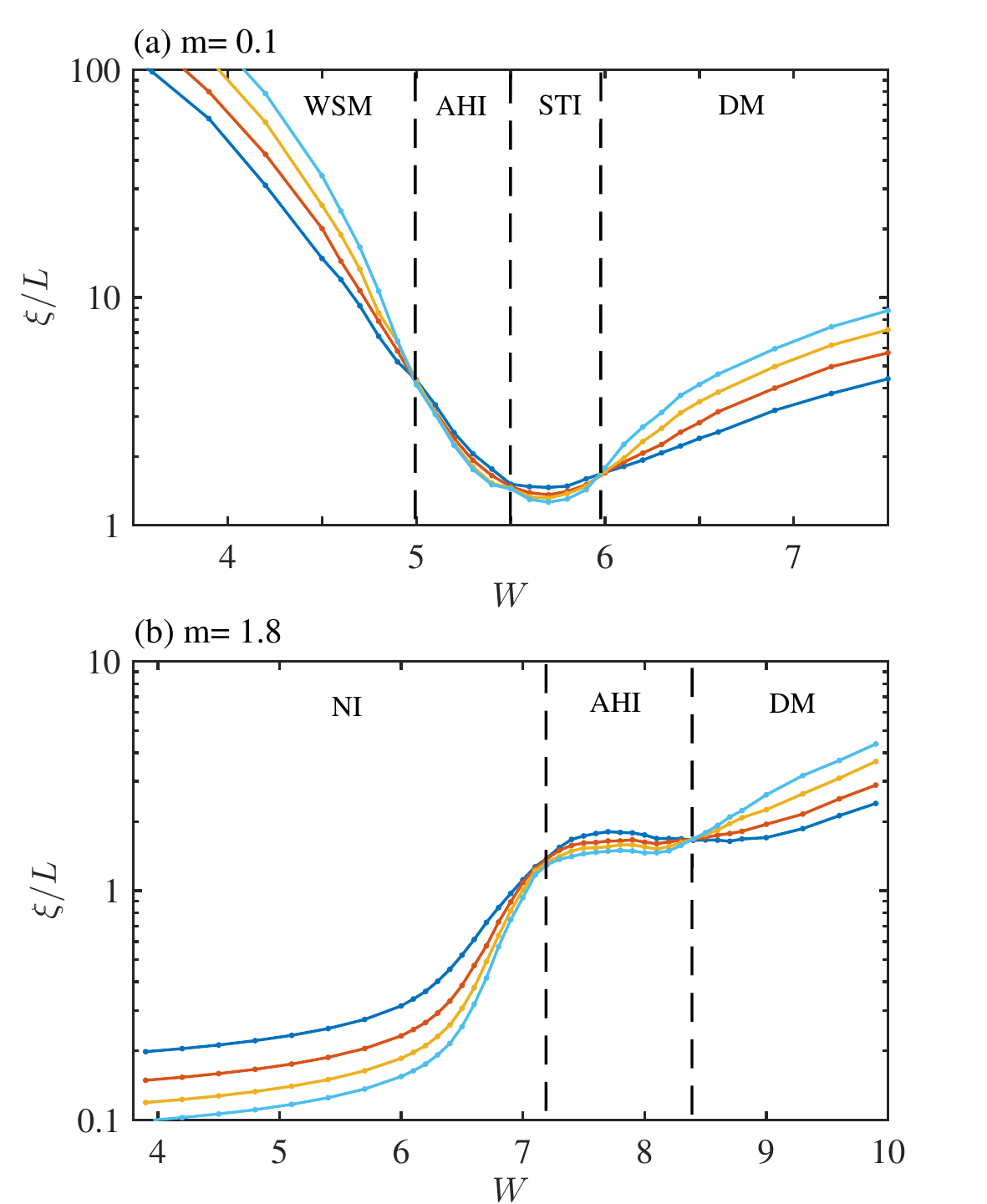}
\caption{\label{fig:4b_Lyp_comp} 
The localization length as a function of disorder strength in the WSM generated by TRS-broken 3D Dirac model with  $b_3=0.6$ for $m=0.1$ (a) and $m=1.8$ (b). The dashed vertical lines represent phase boundaries. The colors show different cross-section sizes $L=6$(blue), $8$(red), $10$(orange), and $12$(cyan). In panel (a), we should note that there is a hint of the scale-invariant point between AHI-STI.}
\end{figure}

Figure~\ref{fig:AHE_fourband} presents the phase diagram of this model where $\sigma_{xy}$ is shown as the background color. 
The solid lines are phase boundaries determined by the scale invariant points of the localization length. 
Typical results for the Hall conductivity and localization length as a function of disorder strength for four $m$ points are also shown in Fig.~\ref{fig:4b_comp}.
For comparison, the phase diagram of $\mathbb{Z}_2$ STI in the absence of the Zeeman field has been studied previously~\cite{Ryu_Nomura_3DTI,Tomi_dis_TI,Sbierski_3DTI,Tomi_DOS}. The apparent shift of the phase boundaries to the right (toward larger $m$) is controlled by the higher order terms in momenta in the effective theory Eq.~(\ref{eq:Dirac_eff}). This has also been reported in the previous studies and leads to  the so-called disorder-induced 3D TI phase~\cite{Ryu_Nomura_3DTI,Tomi_dis_TI}. However, the phase boundary between the STI and NI, which was originally a line~\cite{Ryu_Nomura_3DTI,Tomi_dis_TI}, now broadens into a semi-metallic phase. Notice that the top left region of the phase diagram in Fig.~\ref{fig:AHE_fourband} with negative $\sigma_{xy}$ corresponds to the tail of another WSM phase which originally forms in the mass range $|m+2r|<b_3$ between the STI and a weak TI in the clean limit and is shifted to the right in the diffusive limit (see also Fig.~\ref{fig:4b_comp}(a)), similar to the WSM ($|m|<b_3$) that is our focus here. 

Let us remark that the label STI may be a bit imprecise in presence of TRS breaking Zeeman term. However, in what we call STI, the Zeeman term $b_3$ is weak compared to the bulk gap $m$ so that the bulk wave-functions are not affected much and can be adiabatically connected to those of the time-reversal symmetric STI in absence of the Zeeman term. This implies that the bulk STI (in presence of weak Zeeman) has a zero Hall conductance. However, the surface states are gapped by the Zeeman term and this leads to surface AHE. In Fig.~\ref{fig:4b_comp}(a), we impose periodic BCs along the $z$-direction and the surface contribution is absent hence leading to $\sigma_{xy}=0$. For consistency we checked that open BCs in the $z$-direction yields the quantized AHE ($\sigma_{xy}\cdot L_z=1$) due to the gapped Dirac surface states, although we do not present the figure here. The latter set-up is very similar to a time-reversal symmetric STI along with TRS breaking ferromagnetic layers on the top/bottom surfaces.
Consequently, the STI phase is clearly different from NI, where there is neither bulk nor surface Hall conductance. In the clean limit it can also be thought of as a TI protected by inversion symmetry. The key feature is that the system still exhibits the characteristic topological magnetoelectric effect (quantized surface Hall conductance). 

In addition, we surprisingly find that the semi-metallic phase is interrupted by an insulating phase before turning into the DM. This insulating phase -which we call a disorder induced anomalous Hall insulator (AHI)- must be contrasted from the NI and STI as it has a non-zero \emph{bulk} Hall conductivity (the NI does not have a Hall conductivity and the STI, at most, has a surface Hall conductivity). The AHI emerges in the strong disorder regime as a result of localizing the Weyl nodes without annihilating each other; i.e., in the effective BZ picture, the nodes are far away from each other and localize on their own. Therefore, the AHI phase with non-quantized 3D Hall conductivity (per layer) $\sigma_{xy}$ (as opposed to the weak-disorder limit of the 3D CI studied above where $\sigma_{xy}=1$ per layer) is an exotic insulating  disordered phase which cannot be realized in systems with translational symmetry (which implies the $\sigma_{xy}$ is quantized to be the same multiple of $e^2/h$ for each layer\cite{Halperin1987,*Halperin1992,PhysRevB.41.11417}).  
It is important to note that from our numerics that if we view the AHI as a quasi-2D insulator, it  could be interpreted as a Hall liquid with $\sigma^{2D}_{xy}\neq 0$, but there is another possibility that the AHI phase in the thermodynamic limit converges to the ``Hall insulator'' state, originally introduced in Ref.~\onlinecite{Kivelson1992}, and also experimentally observed recently~\cite{Kapitulnik} at the magnetic-field-driven superconductor to insulator transition of amorphous indium oxide, where $\sigma_{xx}^{2D}\to 0$ and $\sigma_{xy}^{2D}\to 0$, in such a way that $\sigma_{xy}^{2D}\propto (\sigma_{xx}^{2D})^2$ so that $\rho_{xy}^{2D}$ becomes finite. In order to fully characterize the AHI phase one needs to compute $\sigma_{xx}^{2D}$ and $\sigma_{xy}^{2D}$ in the thermodynamic limit at low temperatures and we cannot rule out either possibility based on our current numerical data.

Based on our data, the physical difference between STI and AHI is the existence of chiral edge modes (in the xy plane circulating around the z axis) in the latter when the boundary condition is periodic along the z direction. In other words, the chiral edge modes are descendants of the chiral Fermi arcs in the Weyl semi-metallic phase that the AHI is originated from. 
We anticipate that in order to transition from the AHI phase to either of STI and NI phases,  the system must undergo a (mobility) gap closing (scale-invariant critical point) where the extended bulk states emerge, leading to a sharp change in the topological response (Hall conductivity) in these insulating phases. Our localization length analysis suggests such a behavior at the NI-AHI and STI-AHI phase boundaries (see Fig.~\ref{fig:4b_Lyp_comp}). As mentioned in the caption of Fig.~\ref{fig:AHE_fourband}, we checked that the AHI phase occupies a closed region of the phase diagram by finding the intersection between the converging phase boundaries of the DM-AHI transition and the NI-AHI transition. 

The last remark we want to make is about the direction in which the Weyl nodes are shifted by weak disorder. Our finding based on $\sigma_{xy}$ implies that for $m>m_c$ (defined above)
the Weyl nodes tend to move away from one another leading to an enhancement of the AHE (Fig.~\ref{fig:4b_comp}(c),(d)); while for $m < m_c$ the Weyl nodes are moved closer together so that they annihilate at finite disorder (Fig.~\ref{fig:4b_comp}(b)).  In both cases, the layers with non-zero Chern number are located between two Weyl nodes, however, the nodes can move either towards or further from each other. This is different from the multi-layer CI realizations in previous sections where disorder always pushes the nodes apart to create a wider topological region (Fig.~\ref{fig:motion}(a)). We can understand this for 3D Dirac model by realizing that the coefficient of the quadratic term $m_\perp$ in Eq.~(\ref{eq:quad}) changes sign as $k_0$ crosses the critical value $k_{c}=\sin^{-1}(b_3/t)$ corresponding to the critical mass $m_c$. This causes the mass renormalization (Eq.~(\ref{eq:CImass_ren})) to pick up the opposite sign depending on whether $m$ is larger or smaller than $m_c$: in the former, nodes are moved apart (similar to previous sections) whereas in the latter, nodes are pushed towards one another.
Thus, the movement direction of Weyl nodes  in this case coincides as well with SCBA analysis as long as this complication is taken into account~\cite{Groth_SCBA}.

\section{\label{sec:discussion} Discussion}

In conclusion, we have studied three different lattice model realizations of Weyl semimetals in the presence of an on-site disorder potential by calculating the localization length and the Hall conductivity. The first two models are based on stacking layers of a 2D Chern insulator with Chern numbers one and two respectively, whereas the third model is obtained by breaking the TRS in a 3D four band Dirac model. We have found that in a WSM phase, weak disorder may effectively act to cause the Weyl nodes to move inside the BZ. The movement direction is not determined by the position of the Weyl nodes, e.g., whether they are close to the origin or boundaries of the BZ (as opposed to Ref.~\onlinecite{CZC}). Instead their movement is actually determined by the Chern number of the 2D (momentum-space) layers in the clean limit where the layered structure of the 3D BZ can be applied. In short, disorder forces the Weyl nodes to expand into the region with zero Chern number instead of the region with non-zero Chern number. We have successfully applied this idea to interpret all our results. Moreover, we have also shown that the SCBA analysis is consistent with this picture in the case of  WSMs with two Weyl nodes. We have seen however, in the case of a WSM with four nodes where all layers carry non-zero Chern number, the SCBA formula predicts no motion for the Weyl nodes although our data clearly shows a different phenomenon. Remarkably, this effect can be explained with the aid of  the phase diagram of a single disordered 2D Chern insulator, which implies that layers with opposite Chern numbers in the vicinity of Weyl nodes hybridize and start to lose their Chern number. Nevertheless, an explicit perturbative understanding of this trend remains an open issue.

We have also observed that weak disorder may nucleate Weyl nodes within the BZ of insulating phases near the WSM phases hence endowing them a finite Hall conductivity; or oppositely, weak disorder may annihilate Weyl nodes in the WSM phases near insulating phases leading to a TRS broken insulator adiabatically connected to a weak or strong topological insulator.
We have found that both types of transitions are consistent with the predictions of SCBA.

In addition, we should note that our AHE response data in the WSM phase is consistent with the recent analytical field-theoretic results~\cite{Burkov_rev,Altland_AHE_2015}, once the underlying assumptions of their calculations are taken into account. These studies have shown that the AHE response of a WSM is robust against weak disorder and always given by the separation of the Weyl nodes (which is an intrinsic property of the WSM). As a result, they found that $\sigma_{xy}$ remains intact even in the diffusive limit. These results rely on two important assumptions: first, the higher order (quadratic at lowest) corrections to the band dispersion are neglected and second, the non-perturbative localization effects which ultimately lead to the Anderson localization are not included. The first assumption implies that the distance between Weyl nodes is not renormalized by the disorder and the second assumption prevents any process causing $\sigma_{xy}$ to decrease. However, these two assumptions cannot be made in our numerical calculations and hence the apparent difference between our numerical analysis and the continuum field-theoretic calculations are unavoidable. Nevertheless, as we show here, the observed trend of AHE in the WSM can be related to the renormalized distance (rather than the original distance in the clean-limit) between two Weyl nodes and the fact that AHE in the diffusive limit decays to zero slowly is consistent with the coexistence of diffusive transport and AHE as argued in the continuum limit calculations. 

Finally, our localization length data signals a surprising effect that is special to the 3D Dirac model and was not found in the multi-layer systems: the WSM region of the phase diagram is separated from the diffusive metal through an emergent Hall insulator with a non-quantized (per layer) bulk Hall conductivity. Although the existence of a disordered anomalous Hall insulator with a non-quantized bulk Hall conductivity is not violated by any physical principles, it is not really clear to us that why this phase only appears in the 3D Dirac model. 
One interesting direction is to possibly characterize this phase further and check if it survives in the thermodynamic limit.

\acknowledgements
We would like to acknowledge insightful discussions with B. A. Bernevig, D. Huse, S. T. Ramamurthy, and S. Ryu. 
 In particular, we would like to thank S. A. Kivelson for pointing out an alternate interpretation for the anomalous Hall insulator.
TLH is supported by ONR YIP Award N00014-15-1-2383. 
Computational resources were provided by the Taub campus cluster at the University of Illinois at Urbana-Champaign.

\begin{appendix}

\section{Disordered Chern Insulator}
\begin{figure}
\includegraphics[scale=0.64]{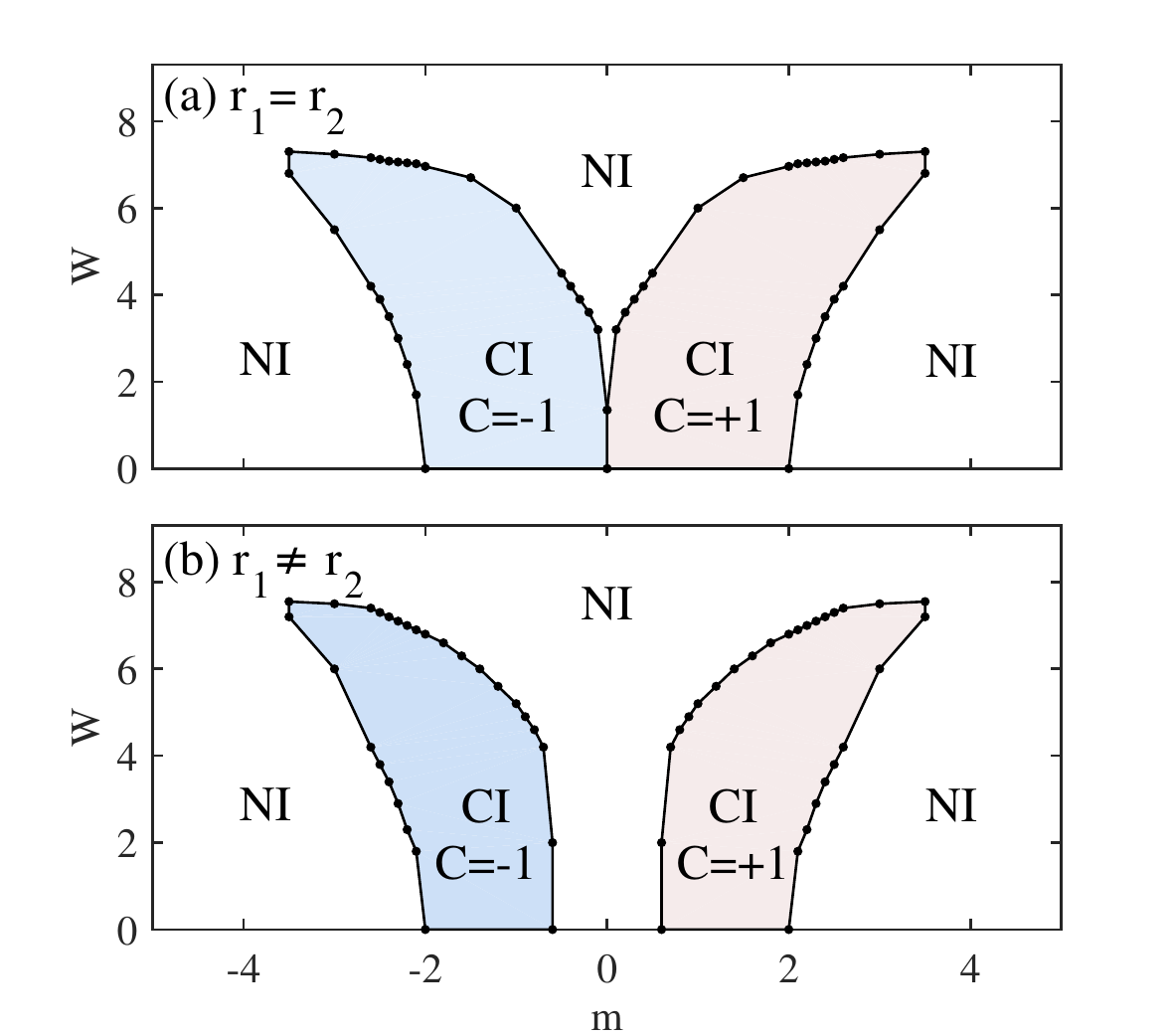}
\caption{\label{fig:2DCI}
Phase diagram of a disordered 2D Chern insulator that serves as a building block of the 3D WSM phase with (a) isotropic ($r_1=r_2=t$) and (b) anisotropic ($r_1=1.3t$, $r_2=0.7t$) Wilson-Dirac mass terms. The chemical potential is set to zero.}
\end{figure}

In this appendix, we use the localization length and Chern number to numerically determine the phase diagram of a disordered 2D Chern insulator model (Fig.~\ref{fig:2DCI}). We use a generic form of the two band model,
\begin{align}
h(\textbf{k})=&  t \sin k_x\ \sigma_1 +t \sin k_y\ \sigma_2 \nonumber \\
&+  ( m-r_1 \cos k_x-r_2 \cos k_y ) \sigma_3. \nonumber
\end{align}
where the Wilson-Dirac mass terms $r_i \cos k_i$ can have anisotropic coefficients. The WSM studied in Sec.~\ref{sec:3DCI} consists of isotropic layers of this model where $r_1=r_2=r$. As Fig.~\ref{fig:2DCI}(a) shows, the clean limit has critical points at $m=-2r$ and $m=2r$ that are bent towards the NI phases with increasing disorder. This is consistent with the SCBA formula (Eq.(\ref{eq:CImass_ren})). Interestingly, the clean critical point at $m=0$, separating the two CI phases, splits to form two phase boundaries and a trivial insulating phase appears in-between them. This behavior is beyond the perturbative SCBA, which gives an exactly vanishing  mass correction in this region. In fact, the SCBA integrals for the continuum limit model are vanishing in this case due to isotropy of the system. From a low-energy perspective,  at $m=0$ the isotropic system carries simultaneous  2D Dirac nodes at $(0,\pi)$ and $(\pi,0)$ in the clean limit. Then, at finite disorder, each of the new phase boundaries must essentially contain one Dirac node to allow for a change in the Chern number by one.

In other words, as we tune $m$ from $+\infty$ to $-\infty$ at a finite disorder (e.g., at $W\gtrsim 2$), we should encounter four critical points to go through the NI-CI-NI-CI-NI sequence of phases. Such a process is possible at finite disorder, since the two band simultaneous crossings $m=0$ are protected by the $C_4$ lattice symmetry which is lifted in the presence of disorder.
For each disorder realization this symmetry is explicitly broken and in principle, one crossing may occur earlier than the other. However, on average, the $C_4$ symmetry is recovered and the crossing at each boundary cannot be attributed to only one node at $(0,\pi)$ or $(\pi,0)$. In fact, we checked that this is the case, by looking at the Fourier transform of the eigenstates (near zero energy) of the disordered Hamiltonian on either phase boundaries. We found that the averaged peak heights at $(0,\pi)$ and $(\pi,0)$ are equal. %We observed that the Fourier transform of the eigenstates (near zero energy) are peaked at $(0,\pi)$ or $(\pi,0)$, but the average heights are equal.

Note that the above situation, where the SCBA vanishes, is specific to the $C_4$ symmetric model. It is easy to see that if one explicitly breaks the $C_4$ symmetry in the clean limit by choosing $r_1$ different from $r_2$, the two band crossings do not occur simultaneously and there will be an intermediate trivial insulator between two CIs for $|m|<|r_1-r_2|$ (Fig.~\ref{fig:2DCI}(b)). Remarkably, the SCBA integral in this case would be non-zero, proportional to the anisotropy $\delta m \propto \pm |r_1-r_2|/ (r_1+r_2)^2$, and have a correct sign consistent with the curvature of the phase boundaries given by the numerics.  Moreover, we observed that the momentum distribution of the energy eigenstates (near zero-energy) on the phase boundaries in the strongly disordered limit ($W>6$) of both the isotropic and anisotropic models have equal weights at $(0,\pi)$ and $(\pi,0)$, meaning that the original difference in the clean limit between these two models is completely lost as a consequence of large scattering between various momentum modes. This is opposite for the phase boundaries at weaker disorder where the anisotropic model has momentum weight dominated by $(0,\pi)$ or $(\pi,0)$ whereas in the isotropic case, as mentioned above, the averaged peak heights of both momenta are the same.

\end{appendix}
%%%%%%%%%%%%%%%%%%%%%%%%%%%%%%%%%%%%%%%%%%%%%%%%
%%%%%%%%%%%%%%%%%%%%%%%%%%%%%%%%%%%%%%%%%%%%%%%%

\bibliography{disWSM_v11}

\end{document}